\documentclass[english,aps,preprint]{revtex4}
\usepackage[T1]{fontenc}
\usepackage{amssymb}
\usepackage[english]{babel} 
\usepackage{amsmath}
\usepackage[draft]{todonotes}
\usepackage{filecontents}

\newcommand{\bra}[1]{\ensuremath{\langle #1|}}
\newcommand{\ket}[1]{\ensuremath{|#1 \rangle}}

\newcommand{\ketbra}[2]{\ensuremath{| #1 \rangle\hspace{-2pt} \langle #2 |}}

\makeatother

\makeatother

\begin{document}
	\title{Coherent Transport of Quantum States 
	\\ by Deep Reinforcement Learning}
	
		\author{Riccardo Porotti}
	\affiliation{Istituto di Fotonica e Nanotecnologie, Consiglio Nazionale delle Ricerche, Piazza Leonardo da Vinci 32, I-20133 Milano, Italy}
	\affiliation{Dipartimento di Fisica ``Aldo Pontremoli'', Universit\`a degli Studi di Milano, via Celoria 16, I-20133 Milano, Italy}
	\author{Dario Tamascelli}
	\affiliation{Quantum Technology Lab, Dipartimento di Fisica ``Aldo Pontremoli'', Universit\`a degli Studi di Milano, via Celoria 16, I-20133 Milano, Italy}
	\author{Marcello Restelli}
	\affiliation{Politecnico di Milano, Piazza Leonardo da Vinci 32, I-20133 Milano, Italy}
	\author{Enrico Prati}
	\affiliation{Istituto di Fotonica e Nanotecnologie, Consiglio Nazionale delle Ricerche, Piazza Leonardo da Vinci 32, I-20133 Milano, Italy}
	\email{enrico.prati@cnr.it}

\begin{abstract}
				Some problems in physics can be handled only after a suitable \textit{ansatz }solution has been guessed. Such method is therefore resilient to generalization, resulting of limited scope.  The coherent transport by adiabatic passage of a quantum state through an array of semiconductor quantum dots provides a par excellence example of such approach, where it is necessary to introduce its so called counter-intuitive control gate ansatz pulse sequence. Instead, deep reinforcement learning technique has proven to be able to solve very complex sequential decision-making problems involving competition between short-term and long-term rewards, despite a lack of prior knowledge. We show that in the above problem deep reinforcement learning discovers control sequences outperforming the \textit{ansatz} counter-intuitive sequence. Even more interesting, it discovers novel strategies when realistic disturbances affect the ideal system, with better speed and fidelity when energy detuning between the ground states of quantum dots or dephasing are added to the master equation, also mitigating the effects of losses. This method enables online update of realistic systems as the policy convergence is boosted by exploiting the prior knowledge when available. Deep reinforcement learning proves effective to control dynamics of quantum states, and more generally it applies whenever an ansatz solution is unknown or insufficient to effectively treat the problem.
	\end{abstract}
	\maketitle
	
	\maketitle
	
	\renewcommand{\baselinestretch}{2}
	
Some problems in physics are solved thanks to the discovery of an ansatz solution, namely a successful test guess, but unfortunately there is no general method to generate one.
	Recently, machine learning has increasingly proved to be a viable tool to model hidden features and effective rules in complex systems.
	Among the classes of machine learning algorithms, deep reinforcement learning (DRL) \cite{sutton1998reinforcement} is providing some of the most spectacular results for its ability to identify strategies for achieving a goal in a complex space of solutions without prior	knowledge \cite{mnih2015human}.  Contrary to supervised learning, which has already been applied to quantum systems, such as the determination of high fidelity gates and to the optimization of quantum memories by
	dynamic decoupling, DRL has been proposed only very recently to maintain a physical system in its equilibrium condition
	\cite{fosel2018reinforcement}.
	To show the power of DRL, we apply it to the problem of coherent
	transport by adiabatic passage (CTAP) of a quantum state encoded by an electron through an array of quantum dots, whose ansatz solution is notoriously called \textit{counterintuitive }because of its not-obvious barrier control gate pulse sequence. During the coherent adiabatic passage, the electron spends no time in the central quantum dot by the simultaneous modulation of the coupling between the dots which suitably drive the trajectory through the Hilbert space \cite{greentree2004coherent,cole2008spatial,greentree2014darkstate,menchon2016spatial}. The system moves from an initial equilibrium condition to a different one represented by populating the last dot of the array.
	By exploiting such \textit{ansatz} solution of pulsing the barrier control gates between the dots in a "reversed order" with respect of what intuition would naturally suggest, the process displays truly quantum mechanical behavior, provided that the array consists of an odd number of dots.
	We already explored separately silicon-based quantum information processing architectures \cite{rotta2016maximum,rotta2017quantum}, including the coherent transport by adiabatic passage of multiple-spin qubits into double quantum dots \cite{ferraro2015coherent}, and heuristic search methods, such as genetic algorithms, to find a universal set of quantum logic gates \cite{ferraro2014effective,michielis2015universal}, and separately the application of deep reinforcement learning to classical systems \cite{bonarini2008batch,tognetti2009batch,castelletti2013multiobjective}. 
	Here we demonstrate that DRL implemented in a compact neural network can, first
	of all, autonomously discover the analogue of the counter-intuitive gate pulse sequence without any prior knowledge, therefore finding a control path in a problem whose solution is far from the equilibrium of the initial conditions. More importantly, such a method allows to outperform the analytical solutions proposed in the past in terms of process speed and when the system deviates from ideal conditions because of detuning between the quantum dots, dephasing and losses. Under such conditions, no analytical approach does exist, to the best of our knowledge.
	Here we exploit the Trust	Region Policy Optimization (TRPO) \cite{schulman2015trust} to handle the CTAP problem. First, we compare the results discovered by the artificial intelligence algorithm with the \textit{ansatz }solution know from the literature. Next, we apply the method to solve the system when the ground states of the quantum dots are detuned, and when the system is perturbed by the interaction with the uncontrollable degrees of freedom of the surrounding environment, resulting in dephasing and loss terms in the master equation describing the system, for which there is no analytical method. 
	Like for the case of the artificial intelligence learning in the classical Atari environment \cite{mnih2015human}, agent here interacts with a QuTIP \cite{johansson2012qutip} simulation providing the environment of the CTAP by implementing the master equation of the system, and exploiting the information retrieved from the feedback in terms of the temporal evolution of the population of the dots. The use of a pre-trained neural network as a starting point to identify the solution of a modified master equation may further reduce the computation time by one order of magnitude. As a further advantages of such approach, a 2-step temporal Bayesian network analysis allows to identify which parameters of the system are more influencing the process. 
	Our investigation indicates that a key factor is the appropriate definition of the reward function that deters the system from occupying the central quantum dot and rewards the occupation of the last quantum dot. 
	
\section{Steering a quantum system far from final equilibrium state}
Reinforcement learning (RL) is a set of techniques used to learn how to behave in sequential decision-making problems when no prior knowledge about the system dynamics is available, or the control problem is too complex for classical optimal-control algorithms. RL methods can be roughly classified into three main categories: value-based, policy-based, and actor-critic methods \cite{sutton1998reinforcement}. Recently, actor-critic methods have proven to be successful in solving complex continuous control problems~\cite{duan2016benchmarking}. \\ 
The idea behind actor-critic methods is to use two parametric models (e.g., neural networks) to represent both the policy (actor) and the value function (critic). The actor decides in each state of the system which action to execute, while the critic learns the value (utility) of taking each action in each state. Following the critic's advice, the actor modifies the parameters of its policy to improve the performance.
Among the many actor-critic methods available in the literature, we selected Trust Region Policy Optimization algorithm (TRPO)~\cite{schulman2015trust} to find an optimal policy of control pulses to achieve CTAP in a linear array of quantum dots. The choice of TRPO is motivated both by its excellent performance on a wide variety of tasks and by the relative simplicity of tuning its hyper-parameters~\cite{schulman2015trust} (see S.I.).
	
	The coherent transport by adiabatic passage is the solid state version of a method developed for Stimulated Raman Adiabatic Passage (STIRAP)\cite{vitanov2001laser,menchon2016spatial}, relevant for instance in those quantum information processing architectures that require to shuttle a qubit from one location to another by paying attention to minimize the information loss during transport. In solid-state quantum devices based on either silicon \cite{maurand2016cmos} or gallium arsenide \cite{bluhm2011dephasing}, the qubit can be encoded, for instance, into spin states of either excess electron(s) or hole(s) in quantum dots \cite{rotta2017quantum}. CTAP was originally developed for single-electron states in single-occupied quantum dots, but it can be also extended to more complex spin states, such as for hybrid qubits based on triplets of spin.\cite{ferraro2015coherent} If one imagines to employ, for instance, an array of dopants in silicon \cite{prati2012anderson}, a reasonable inter-dopant spacing is of the order of about 20 nm and hopping time of 100 ps \cite{hollenberg2006two}. The adiabatic passage needs control pulses with a lower bandwidth of an order of magnitude or two with respect to the hopping time, which can be managed by conventional electronics \cite{homulle2017reconfigurable}.
	To demonstrate the exploitation of deep reinforcement learning, we start by the simplest case of CTAP across a chain of three identical quantum dots.
	The deep reinforcement learning architecture is depicted in Figure 1a. The simulation of the physical system that supports the CTAP consists in the environment E that receives as input the updated values of the parameters (in our case the coupling terms $\Omega_{i,i+1}$ with $i=1,2$ between the adjacent dots \textit{i}${}^{th}$ and (\textit{i}+1)${}^{th}$) that reflect the action on the control gates as calculated by the agent A according to the policy $\pi$. In turn, the environment E computes the new system state (here expressed in terms of the density matrix of the triple quantum dot device) and provides feedback to agent A. Agent A calculates the next input parameters after evaluating the effects of the previous input according to a reward function $r_t$, which is expressed in terms of the system state.
	\\
	More explicitly, the ground state of each quantum dot is tuned with respect to the others by
	external top metal gates (not shown for simplicity in the sketch of the environment E in Figure 1a),while the coupling between two neighboring quantum dots is in turn controlled by additional barrier control gates. The idealized physical system is prepared so that the ground states of the three quantum dots have the same energy. As the reference energy is arbitrary, we can without any loss of generality set $E_1=E_2=E_3=0$. The Hamiltonian therefore reads:
	\begin{equation} \label{eq:ham}
	H=\left( \begin{array}{ccc}
	0 & -\Omega_{12} & 0 \\ 
	-\Omega_{12} & 0 & -\Omega_{23} \\ 
	0 & -\Omega_{23} & 0 \end{array}
	\right). 
	\end{equation}
	One of the three eigenstates is special in that it is expressed as a function of the state of the first and third dots only and reads
	\begin{equation}
	\left|D_0\right\rangle ={\mathrm{cos} \theta_1\ \left|1\right\rangle
		-sin\theta_1|3\rangle, \
	}
	\end{equation} 
	where $\theta_1=\arctan(\Omega_{12}/\Omega_{23}$). A suitable time evolution of the values of the coupling terms $\Omega_{12}(t)$ and $\Omega_{23}(t)$ between the dots allows to transform $|D_0\rangle $ from $\left|1\right\rangle$ at $t=0$  to $\left|3\right\rangle$ at $t=t_{max}$. If the Hamiltonian is prepared in$\left|D_0\right\rangle$ at $t=0$, it will remain in the same eigenstate if the adiabaticity criterion is met, that is $|E_0-E_{\pm }|\gg |\langle D_0|D_{\pm}\rangle|$ where $D_{\pm }(t)$ eigenstates are  explicitly expressed in the S.I.. The effectiveness of the pulse sequence of the barrier control gate, which is reflected on the coupling terms $\Omega_{i,i+1}$, is addressed by taking $F(t)=\rho_{33}(t)$ as the fidelity and by maximizing it while the population $\rho_{22}(t)$ is kept constant at zero. 
	Time evolution is governed by a master equation that involves the density matrix $\rho$, namely $\dot{\rho}=-\frac{i}{\hslash }[H,\rho]$. Notoriously a suitable shaping of $\Omega_{12}(t)$ and $\Omega_{23}(t)$ as Gaussian pulses can achieve a coherent transport with high fidelity, if $t_{max}\gtrsim  \frac{10\pi}{\Omega_{max}}$. The remarking fact is that the two pulses must be applied in the so-called \textit{counter-intuitive} sequence, with the meaning that the first gate controlling $\Omega_{12}(t)$ is operated as second in time, while the second gate acting on $\Omega_{23}(t)$ as first. Such pulse sequence drives the occupation of the first dot ${\rho}_{11}(t)$ to zero and that of the last dot ${\rho}_{33}(t)$ to 1, while maintaining the central dot empty (${\rho}_{22}(t)=0$) \cite{greentree2004coherent}. 
	It is worth mentioning that recently a different \textit{ansatz }combination of pulse shapes has been proposed to speed up such process \cite{ban2018fast}. Generally speaking, there is no analytical method to optimize the pulses, so further improvements are based only on still undiscovered ideas. Here is where the power of the deep reinforcement learning comes into play. Thanks to a robust definition of the \textit{reward function }that allows the agent to judge its performance, the neural network evolves in order to obtain the best temporal evolution of the coupling parameters $\Omega_{i,i+1}$, so to ensure that the wanted effects on the electron population across the device evolved over time. The definition of the best reward function is certainly the most delicate choice in the whole model. The key features of the reward can be summarized by two properties, namely its generality, so it should not contain specific information on the characteristics of the two pulses, and its expression as a function of the desired final state, which in our case consists of maximizing $\rho_{33}\ $at the end of the temporal evolution. The rewards functions used in this research are fully accounted for in the S.I.. The reward function used in most simulations is
	\begin{equation}\label{eq:reward}
	r_{t}=\alpha(-1+\rho_{33}\left(t\right)-\rho_{22}\left(t\right))-e^{\beta{\cdot \rho}_{22}(t)},
	\end{equation}
	with $\alpha, \beta>0$. The sum of the first three terms is non-positive at each time step, so the agent will try to bring it to 0 by minimizing $\rho_{22}$ and by maximizing $\rho_{33}$. Subtracting $e^{\beta \rho_{22}}$ (e.g., punishing electronic occupation in the site 2) improves the convergence of the learning. Furthermore, in some specific cases, we stop the agent at an intermediate episode if $\rho_{33}$ is greater than an arbitrary threshold $\rho^{th}_{33}$ for a certain number of time step. This choice can help to find fast pulses that achieve high-fidelity quantum information transport, at the cost of a higher $\rho^{max}_{22}$ with respect to the analytic pulses (more details are given in S.I.).\\ 
	Figure 3 shows the best results achieved by the agent at various epochs. As the agent is left free of spanning the whole coupling range at each step, we smoothed the output values $\Omega_{i,i+1}$ and we run the simulation for the smoothed values, with negligible differences from the simulation calculated by using the original output values. The smoothing is done to obtain a more regular shape of the pulses by a spline interpolation. The three occupations $\rho_{ii}$, $i=(1,2,3)$ shown in Figure 3 refer to the smoothed pulse sequences. At the very beginning, the untrained agent tries random values with apparently small success, as the occupation oscillates between the first and the last dot during time. It is worth noting that despite such fluctuation, the agent, thanks to the reward function, learns very quickly after only 45 epochs to keep the occupation of the central dot always below 0.5. After about 2000 epochs, the agent learns to stabilize the occupation of the first and last dot, while	keeping the central dot empty. After about twice the time, the reversing of the population of the first and last dot happens, even if they do not yet achieve the extremal values. Finally, after about four times more, the agent achieves a high-fidelity CTAP. Notice that the pulse sequence reminds the ansatz Gaussian counterintuitive pulse sequence as the second gate acting on $\Omega_{23}$ is operated first, but the shape of the two pulses differs. It is remarkable that the reward function implicitly asks to achieve the result as quickly as possible, resulting in a pulse sequence that is significantly faster than the analytical Gaussian case, and comparable to the recent proposal of Ref. \cite{ban2018fast}. The high point is that the agent achieves such results irrespectively on the actual terms of the Hamiltonian contributing to the master equation. Therefore, DRL can be applied straightforward to more complex cases for which there is no generalization of the \textit{ansatz} solutions found for the ideal case, to which we devote the next section.
	
\section{Deep reinforcement learning to overcome disturbances}
We turn now the attention to the behavior of our learning strategy applied to the non-ideal scenario in which the realistic conditions typical of semiconductor quantum dots are considered. In particular, we discuss the results produced by DRL when the dot array is affected by detuning between the energy gap of the dots, dephasing and losses. Such three effects do exist, to different degrees, in any practical attempt to implement CTAP of the electron spin in quantum dots: the first one is typically due to manufacturing defects \cite{jehl2011mass}, while the last two emerge from the interaction of the dots with the surrounding environment \cite{breuer02,clement2010one} involving charge and spin fluctuations \cite{pierre2009background,kuhlmann2013charge} and noise on the magnetic field \cite{prati2013quantum}. Under such kind of disturbances, neither analytical nor \textit{ansatz }solutions are available. On the other side, the robustness and generality of the reinforcement-learning approach can be exploited naturally as, from the point of view of the algorithm, it does not differ from the ideal case discussed above. As we could determine the pulse sequence only by adding a second hidden layer H2 of the neural network, we refer from now on to deep reinforcement learning.\\
Let us consider a system of $N=3$ quantum dots with different energy $E_i, i={1,2,3}$. We indicate by $\Delta_{ij}=E_j-E_i$, so that the Hamiltonian (\ref{eq:ham}) can be written, without loss of generality, as
\begin{equation} \label{eq:hamDetuning}
H=\left( \begin{array}{ccc}
0 & -\Omega_{12} & 0 \\ 
-\Omega_{12} & \Delta_{12} & -\Omega_{23} \\ 
0 & -\Omega_{23} & \Delta_{13} \end{array}
\right).
\end{equation}
Figure 3b refers to the particular choice $\Delta_{12}=\Delta_{23}=0.15 \Omega_{max}$ (a full 2D scan of both $\Delta_{12}$ and $\Delta_{13}$ is shown in the S. I.). In this case DRL finds a solution that induces a significantly faster transfer than the one
obtained with standard counter-intuitive Gaussian pulses. Moreover, the latters are not even able to achieve an acceptable fidelity. As shown in the S. I., such speed is a typical feature of the pulse sequences determined by DRL. \\
Besides energy detuning between the quantum dots, in a real implementation, the dots interact with the surrounding environment. Since the microscopic details of such interaction are unknown, its effects are taken into account through an effective master equation. A master equation of Lindblad type with time-independent rates is adopted to grant sufficient generality while keeping a simple expression. To show the ability of the DRL to mitigate disturbances, we consider, in particular, two major environmental effects consisting of decoherence and losses respectively. The first corresponds to a randomization of relative phases of the electron states in the quantum dots, which results in a cancellation of the coherence terms, i.e., the off-diagonal elements of the density matrix in the position basis. The losses, instead, model the erasure of the quantum information carried by the electron/hole moving along the dots. In fact, while the carrier itself cannot be reabsorbed, the quantum information, here encoded as a spin state, can be changed by the interaction with some surrounding spin or by the noise on the magnetic field. In the presence of dephasing and losses, the system's master equation is of Lindblad type. Its general form is
\begin{equation}
\label{eq:lindblad}
\dot{\rho}= -\frac{i}{\hbar} \left[H,\rho \right ]+\sum_n \Gamma_n \left( A_n \rho A_n^\dagger-\frac{1}{2} \left \{A_n^\dagger A_n,\rho \right\}\right),
\end{equation}
where $A_k$ are the Lindblad operators, $\Gamma_k$ the associated rates, and $\{A,B\} =AB+BA $ is
the anticommutator. \\
We first considered each single dot affected by dephasing  and assumed equal the dephasing rate $\Gamma_d$ for each dot. The master equation can be rewritten as
\begin{equation}
\dot{\rho}=-\frac{i}{\hbar}\left[H,\rho \right]+\Gamma_{d} \left[ \rho -\mbox{diag}(\rho) \right].
\end{equation}
Figure 3c shows an example of the pulses determined by DRL and the corresponding dynamics for the choice $\Gamma_{d}=0.01 \Omega_{max}$ . The CTAP is achieved with Gaussian pulses in a time $t\approx 10\pi/\Omega_{max}$ at the cost of a significant occupation of the central dot (inset). The advantage brought by the DRL agent with respect to Gaussian pulses is manifest: DRL (solid lines) realizes the population transfer in about half of the time required when the analytic pulses (dashed lines) are employed. The occupation of the central dot is less than the one achieved by Gaussian pulses.\\
We now consider the inclusion of a loss term in the master equation. The loss refers to the information carried by the electron encoded by the spin. While the loss of an electron is a very rare event in quantum dots, requiring the sudden capture of a slow trap \cite{prati2008giant,prati2010measuring}, the presence of any random magnetic field and spin fluctuation around the quantum dots can modify the state of the transferred spin. Losses are described by the operators $\Gamma_{l} \ketbra{0}{k}, k=1,2,3$, where $\Gamma_{l}$ is the loss rate, modeling the transfer of the amplitude from the $i$-th dot to an auxiliary \emph{vacuum} state $\ket{0}$. Figure 3d shows the superior performance of DRL versus analytic Gaussian pulses. 
Because of the reward function, DRL minimizes transfer time as to minimize the effect of losses.	
As a further generalization, we applied DRL to the passage through more than one intervening dot, an extension called straddling CTAP scheme (or SCTAP) as from Refs. \cite{malinovsky1997simple,greentree2004coherent}. To exemplify a generalization to odd $N>3$, we set $N=5$ according to the sketch depicted in Figure 4a, and no disturbance is considered for simplicity. For 5-state transfer, like 3-state and for any other odd number of quantum dots, there is one state at zero energy. The realization of the straddling tunneling sequence is achieved by augmenting the original pulse sequence by the straddling pulses, identical for all intervening tunneling rates $\Omega_{middle}$. The straddling pulse involves the second and third control gates, as shown in Figure 4a. The pulse sequence discovered by DRL (Figure 4b) is significantly different from the known sequence based on Gaussian pulses (Figure 4a, right) and it operates the population transfer in about one third of the time by displaying only some occupation in the central dot. It is remarkable that the reward function (see S.I.) achieves such a successful transfer without any requirement on the occupation of the intermediate dots $\rho_{i,i}$ with (i=2,3,4) as it only rewards occupation of the last dot.

\section{Analysis of relevant variables within the deep reinforcement learning framework}
The advantages of employing DRL with respect of an ansatz solution to solve the physical problem is further increased by the chance of determine which factors are more relevant to the effectiveness of the final solution by the analysis of the neural network. In fact, the employment of DRL algorithm enables the analysis of which state variables are actually relevant to solve the control problems, like that discussed above. To select the variables needed to solve an MDP, we follow the approach presented in~\cite{peters2008natural,peters2010relative,castelletti2011tree}. The idea is that a state variable is useful if it helps to explain either the reward function or the dynamics of the state variables that in turn are necessary to explain the reward function. Otherwise, the state variable is useless and it can be discarded without affecting the final solution. To represent the dependencies between the variables of an MDP, we use a 2-Step Temporal Bayesian Network (2TBN)~\cite{koller2009probabilistic}. In the graph of Figure 5, there are three types of nodes: the Grey diamond node represents the reward function, the circle nodes represent state variables and the squares represent action variables. The nodes are arranged on three vertical layers: the first layer on the left includes the variables at time $t_{i}$, the second layer the state variables at time $t_{i+1}$, and the third layer the node that represents the reward function. If a variable affects the value of another variable, we obtain a direct edge that connects the former to the latter. The weights are quantified in the S.I.. Figure 5 shows the 2TBN estimated from a dataset of 100,000 samples obtained from the standard case of the ideal CTAP. As expected from Eq.~\ref{eq:reward}, the reward function depends only on the values of variables $\rho_{22}$ and $\rho_{33}$. From the 2TBN, it emerges that the dynamics of these two state variables can be fully explained by knowing their value at the previous step, the values of the two action variables $\Omega_{12}$ and $\Omega_{23}$ and the actions taken in the previous time step (stored in the variables $\Omega_{12}'$ and $\Omega_{23}'$). All other state variables do not appear and therefore could be ignored for the computation of the optimal control policy. Such finding matches the expectation from the constraints of the physical model that the coherences are not directly involved in the dynamics and that $\rho_{11}$ is linearly dependent from $\rho_{22}$ and $\rho_{33}$ as the trace of the density matrix is constant. To confirm this finding, the agent has been successfully trained in the standard case of the ideal CTAP by using only the input values ($\rho_{22}$, $\rho_{33}$, $\Omega_{12}$, $\Omega_{23}$). Furthermore, the size of the hidden layer H1 could be reduced in this case from 64 to 4 with the same convergence of the total reward as a function of the episodes, or even faster by reducing the batch size from 20 to 10 (see S.I.). 
In the detuning and dephasing cases, the corresponding 2TBNs are more complex since the dynamics of $\rho_{22}$ and $\rho_{33}$ are affected also by the values of the other elements of the matrix $\rho$. This fact is also corresponding to the changes carried by the Hamiltonian, as trace is not constant when loss are present and because of the direct action on the coherence for the dephasing. 
	
\section{Conclusions}
Neural networks can discover control sequences of a tuneable quantum system starting its time evolution far from its final equilibrium, without any prior knowledge. Contrary to the employment of special ansatz solutions, deep reinforcement learning discovers novel sequences of control operations to achieve a target state, regardless possible deviations from the ideal conditions. Such method is irrespective from having previously discovered \textit{ansatz }solutions and it applies when they are unknown. 
To apply to a practical quantum system known for its counter-intuitiveness, we adopted quantum state transport across arrays of quantum dots, including sources of disturbances such as energy level detuning, dephasing, and loss. In all cases, the solutions found by the agent outperform the known solution in terms of either speed or fidelity or both. Both pre-training of the network and 2TBN analysis - by reducing the computational effort - contribute to speed up the learning process. 
In general, we have shown that neural-network based deep reinforcement learning provides a general tool for controlling physical systems by circumventing the issue of finding ansatz solutions when neither a straightforward method nor a brute force approach are possible. 
	
\section{Acknowledgements}
E. Prati gratefully acknowledges the support of NVIDIA Corporation for the donation of the Titan Xp GPU used for this research.

\section{Methods}
\subsection{Master equation for Hamiltonian including pure dephasing}
Pure dephasing \cite{Taylor2008} is the simplest model that accounts for environment interaction of an otherwise closed system. The corresponding Lindblad operators are diagonal and their form is:
\begin{equation}
L_n=\sqrt{\gamma_n} \ket{n} \bra{n}.
\end{equation}
In the case $\gamma_1=\gamma_2=\gamma_3=\Gamma$, the master equation becomes:
\begin{equation}
\dot{\rho}=-\frac{i}{\hbar}\left[H,\rho \right]+\Gamma \left[ \rho -diag(\rho) \right].
\end{equation}
If the Hamiltonian is time-independent, then
\begin{equation}
\frac{d}{dt} \rho_{ii}=0
\end{equation}
for every $i$, so the electronic population remains unchanged in presence of pure dephasing. This implies \cite{Tempel} that the energy of the system remains unchanged during the evolution, since the environment cannot change it.\\

\subsection{Neural network}
	The neural network consists of a multi-layered feed-forward network of $(N_i,H_1,[H_2,]N_o)$ neurons, where $N_i=11$, as it includes the nine elements of the density matrix $\rho$ and the two coupling terms $\Omega_{ij}$, $N_o=2$ are the two updated values of the coupling terms spanning a continuous range between 0 and an arbitrary value $\Omega_{max}$ chosen according to experimental considerations, and $H_i$ the number of neurons of the $i^{th}$ hidden layer. 

Different setups for the neural network have been employed. Nonethless, all the neural networks were feed-forward and fully connected, with ReLU activation function for the neurons. For the implementation of the neural networks and their training, Tensorforce \cite{schaarschmidt2017tensorforce} has been employed. Being the RL algorithm adopted for this work Trust Region Policy Optimization (TRPO), the hyperparameters have been set as summarized in S.I..
Optimal number of neurons of the hidden layer is peaked around $2^5-2^6$ (see S.I.). 

\subsection{Reward functions}
Various kinds of reward function have been implemented and compared. In general, higher cumulative reward values should correspond to better approximation of an ideal coherent transfer. \\
The difficulty of defining the reward in our case arises from the fact that ideally one wants to keep the maximum value of $\rho_{22}$ small while the maximum value of $\rho_{33}$ (which defines the fidelity) reaches 1.\\
Among those tested, the most effective reward functions $r_t$ have, in general, two characteristics:
\begin{itemize}
	\item A punishing term $-(1-\rho_{33}+\rho_{22})$, so that this subtractive term is $0$ if $\rho_{33}=1$, $\rho_{22}=0$, and is maximum if $\rho_{33}=0$, $\rho_{22}=1$, so an agent will try to minimize the difference between $\rho_{33}$ and 1 and the difference between $\rho_{22}$ and 0.
	\item A punishing term proportional to $e^{\beta \rho_{22}}$, where $\beta>0$. This term exponentially punishes increments of $\rho_{22}$. Typical values for $\beta$ are of the order of units. 
\end{itemize}
Note that such two terms are both non-positive, so a "perfect" simulation can have a maximum of cumulative reward equal to $0$. The two terms are also differentiable, and this can help the agent to converge to a global maximum, because the algorithm used in this work involves derivatives of the reward function. \\
The reward functions used for each case are reported in S.I..
In the case of the SCTAP across an array of $N=5$ quantum dots, the reward function used in Figure 4 of the main text is:
\begin{equation}
\label{reward5QD}
r_t = 1-\rho_{55}
\end{equation} 
Despite its simplicity, the agent was able to keep $\rho_{22}=\rho_{44}=0$ and $\rho_{33}$ close to zero during most of the population transfer.
 
\subsection{Software}
The QuTIP routine has been run in parallel by using GNU Parallel.\cite{tange2011gnu}

\section{Data availability}
The data that support the findings of this study are available from the corresponding author upon reasonable request.

\section{Competing interest}
The authors declare that there are no competing interests

\section{Author contributions}
R. P. developed the simulation and implemented the algorithms, D. T. elaborated on the Hamiltonian framework and the master equation formalism, M. R. operated the reinforcement learning analysis, E. P. conceived the simulated experiment and coordinated the research. All the Authors discussed and contributed to the writing of the manuscript.


\section{Figures}
\begin{figure}
	\centering
	\includegraphics[width=0.95\textwidth]{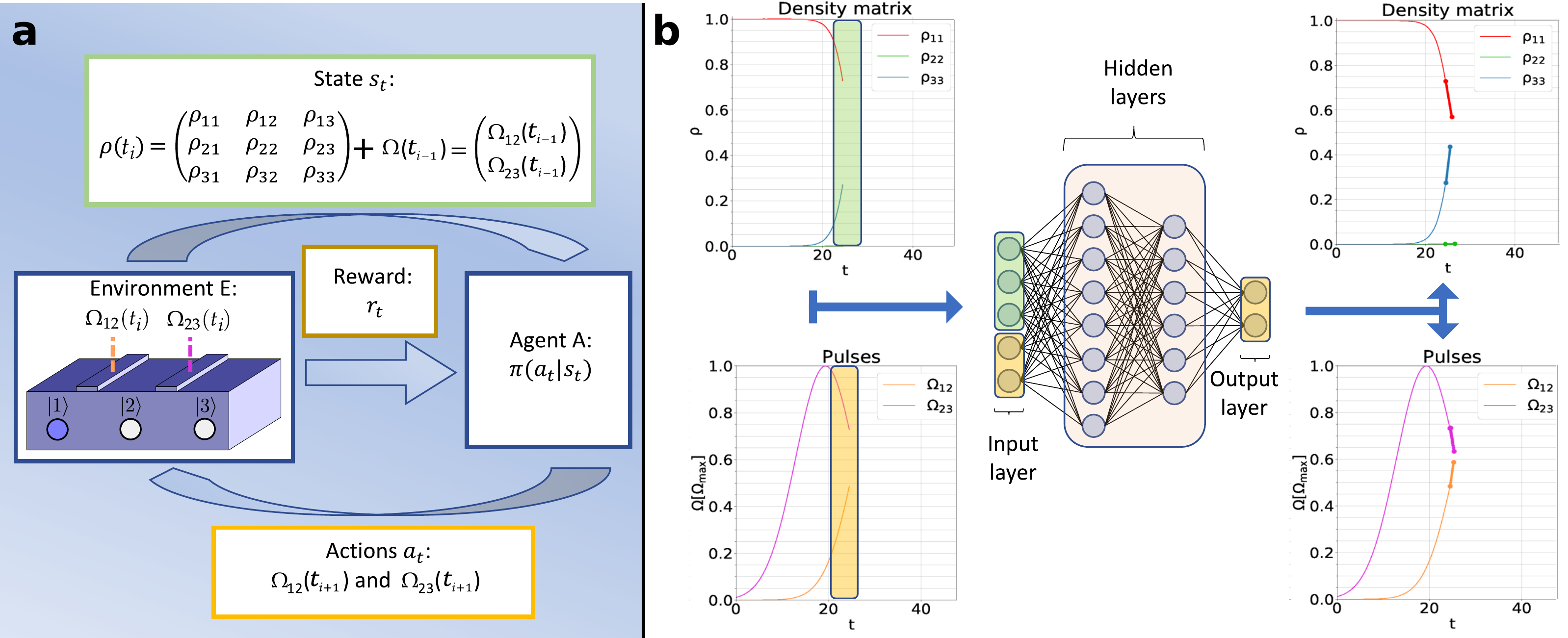}
	\caption{a) Architecture of a deep reinforcement learning architecture. The environment E can be represented by a linear array of quantum dots, having tunneling rates controlled by two gates indicated by $\Omega_{12}$ and $\Omega_{23}$. At each time step, the environment can be modeled by a 3$\times$3 density matrix, that is employed as input observation (the state) by the agent A. In turn, the agent A uses that observation to choose the action in the next time step, by following a policy $\pi(a_t|s_t)$. That action brings E in a new state $\rho_{k+1}=\rho\left(t+\Delta t\right)$. Each action is punished or rewarded with a real-valued reward, indicated by $r_t$. b) The agent A is here represented with a 3-layer neural network, which acts as the policy $p$. The network receives, at each time step, the 3$\times$3=9 real values associated with the density matrix $\rho$ and the 2 values of the gate-control pulses $\Omega_{12}$ and $\Omega_{23}$, for a total of 11 neurons necessary for the input layer. For clarity, in b only the diagonal values of $\rho$ are shown. Then, the agent computes the policy $\pi$ and outputs the values of the two pulses that will be applied to barrier control gates in the next time step (b lower right). Starting and ending points of the highlighted segments can be connected by different functions of time (see Supplementary Information). Finally, the physical simulation of the environment brings the system in a new state by updating the density matrix accordingly, and returns $r_t$ to the agent. When times $t=t_{max}$ the system reaches the final-step ($k=N$) and the simulation is ended.}
	\label{Fig1}
\end{figure}

\begin{figure}
	\centering
	\includegraphics[width=0.95\textwidth]{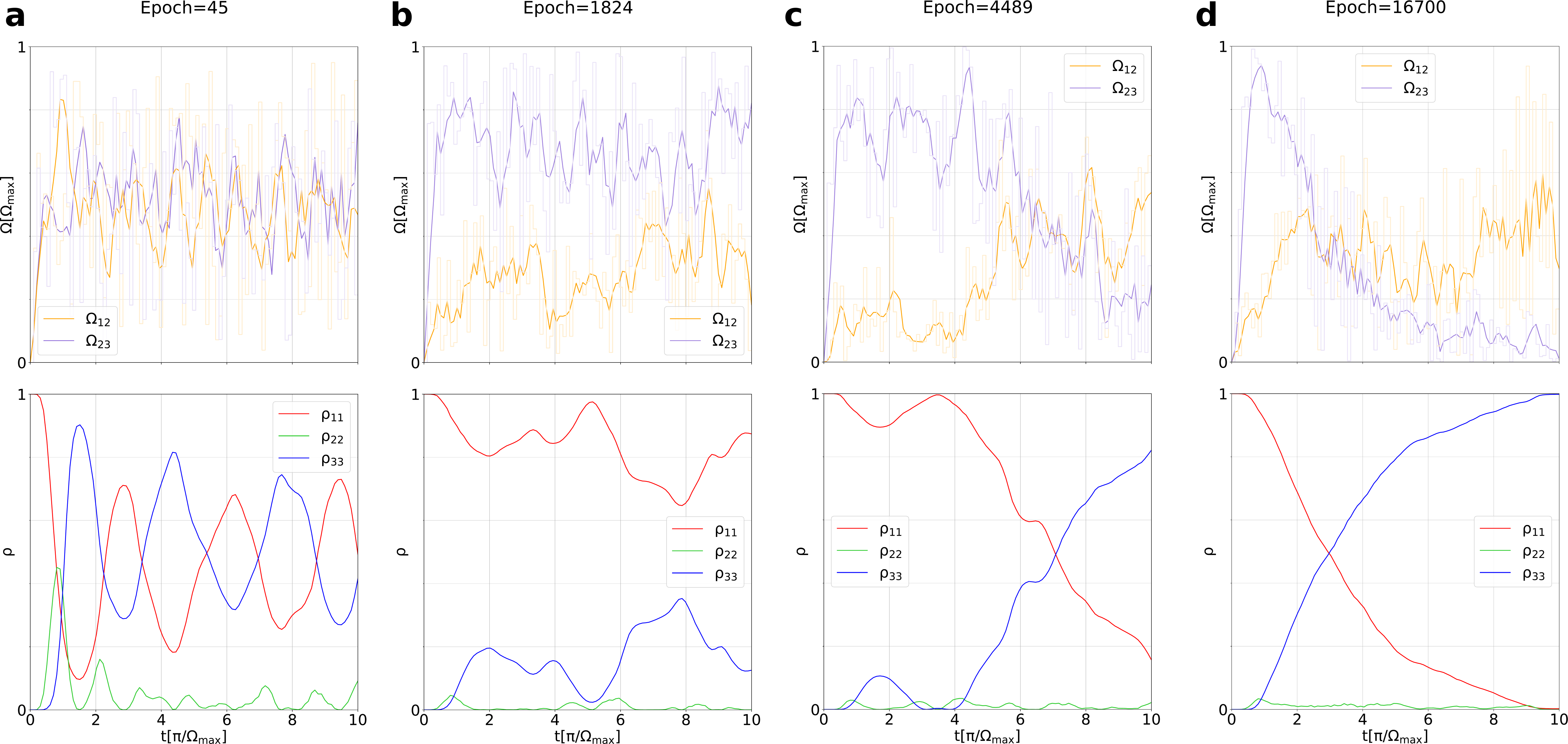}
	\caption{Learning of the agent for various training epochs, for $t_{max}=10\frac{\pi}{\Omega_{max}}$. Every time evolution has been divided in $N=50$ time steps. In every time step $t_i$, the neural network can choose the value of both inter-dot couplings between pair of neighboring quantum dots, which range from $0\ $to $\Omega_{max}$, which are assumed to be constant during each time step. In the top panel of all 4 epochs shown, the faded pulses are the outputs of the DRL agent, while the highlighted lines are a moving average with a 4-sized window. The bottom panels represent the time evolution of the diagonal elements of the density matrix with such averaged pulses applied. We empirically observed that smoothed pulses achieve better results in terms of minimizing $\rho_{22}$ and maximizing $\rho_{33}$.
		Different regimes can be observed: in the first one, the agent tries to explore randomly. Later, the agent learns how to minimize population in the second QD. Finally, it learns how to achieve high population on site 3 at the end of the evolution, without populating the second one.}
	\label{Fig2}
\end{figure}
\begin{figure}
	\centering
	\includegraphics[width=0.65\textwidth]{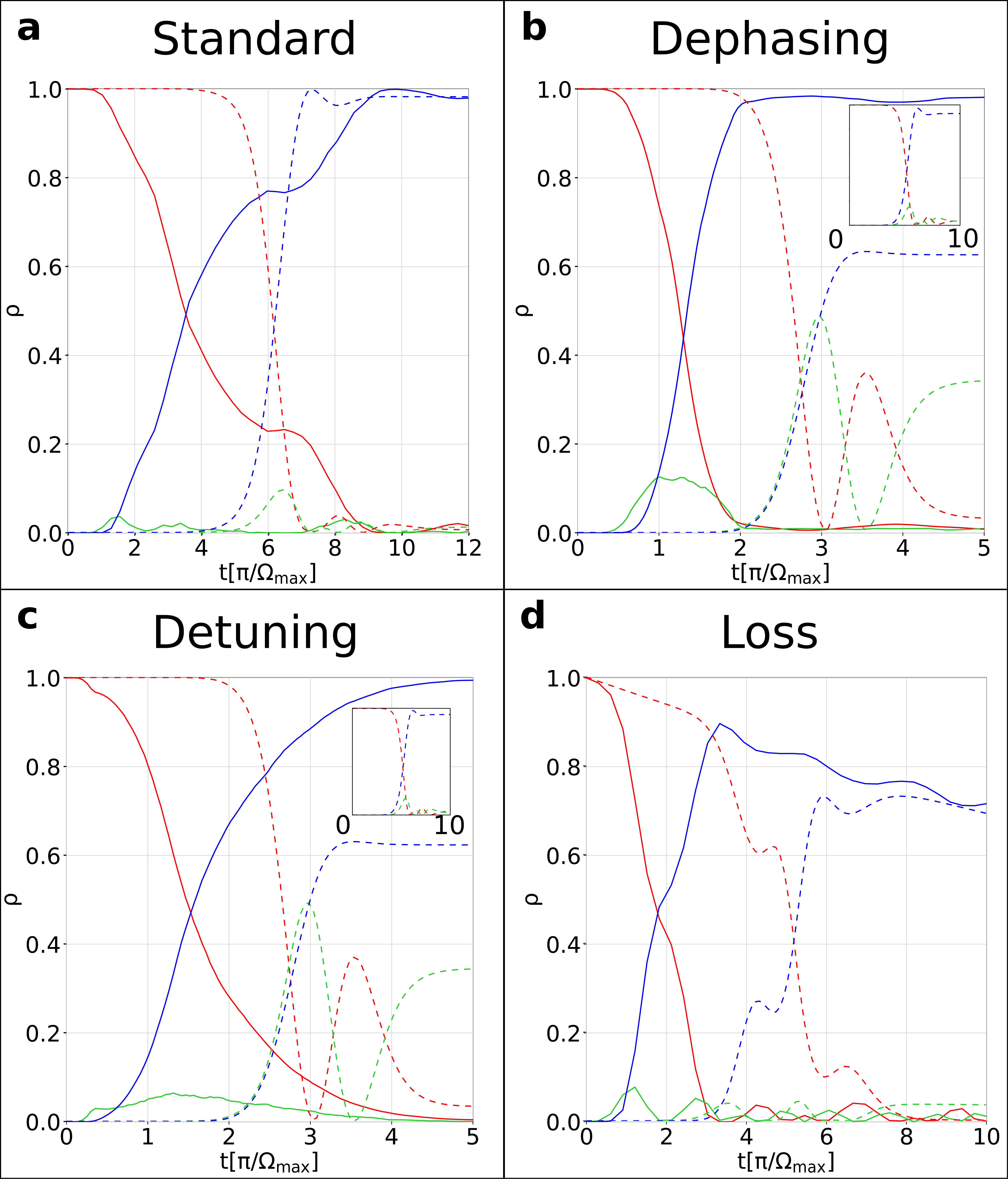}
	\caption{Comparison of CTAP transfer achieved with Gaussian pulses (dotted lines) vs. pulses found by DRL algorithm (solid lines), for different kind of disturbances acting on the system. (a) reference results consisting of simulation in ideal condition with identical eigenvalues in the three quantum dots with $t_{max}=12 \frac{\pi}{\Omega_{max}}$. (b) simulation involving dephasing term in the Hamiltonian, where $\Gamma_d=0.01 \Omega_{max}$ and $t_{max}=5 \frac{\pi}{\Omega_{max}}$. (c) shows a simulation for a detuned system where $\Delta_{1,2}=\Delta_{2,3}=0.15 \Omega_max$ and $t_{max}=5 \frac{\pi}{\Omega_{max}}$. (d) case of arrays of three quantum dots affected by loss, accounted for by an effective term weighted by $\Gamma_l=0.1 \Omega_{max}$ and $t_{max}=10 \frac{\pi}{\Omega_{max}}$. Insets of (b) and (c) show simulations achieved with Gaussian pulses for $t_{max}=10 \frac{\pi}{\Omega_{max}}$. The setup of the neural networks employed were respectively (16,0), (128,64), (64,64) and (128,64), where the first number in parenthesis represents the number of neurons of the first hidden layer H1 of the network and the second one represents the number of neurons of the second hidden layer H2.}
	\label{Fig3}
\end{figure}
\begin{figure}
	\centering
	\includegraphics[width=0.95\textwidth]{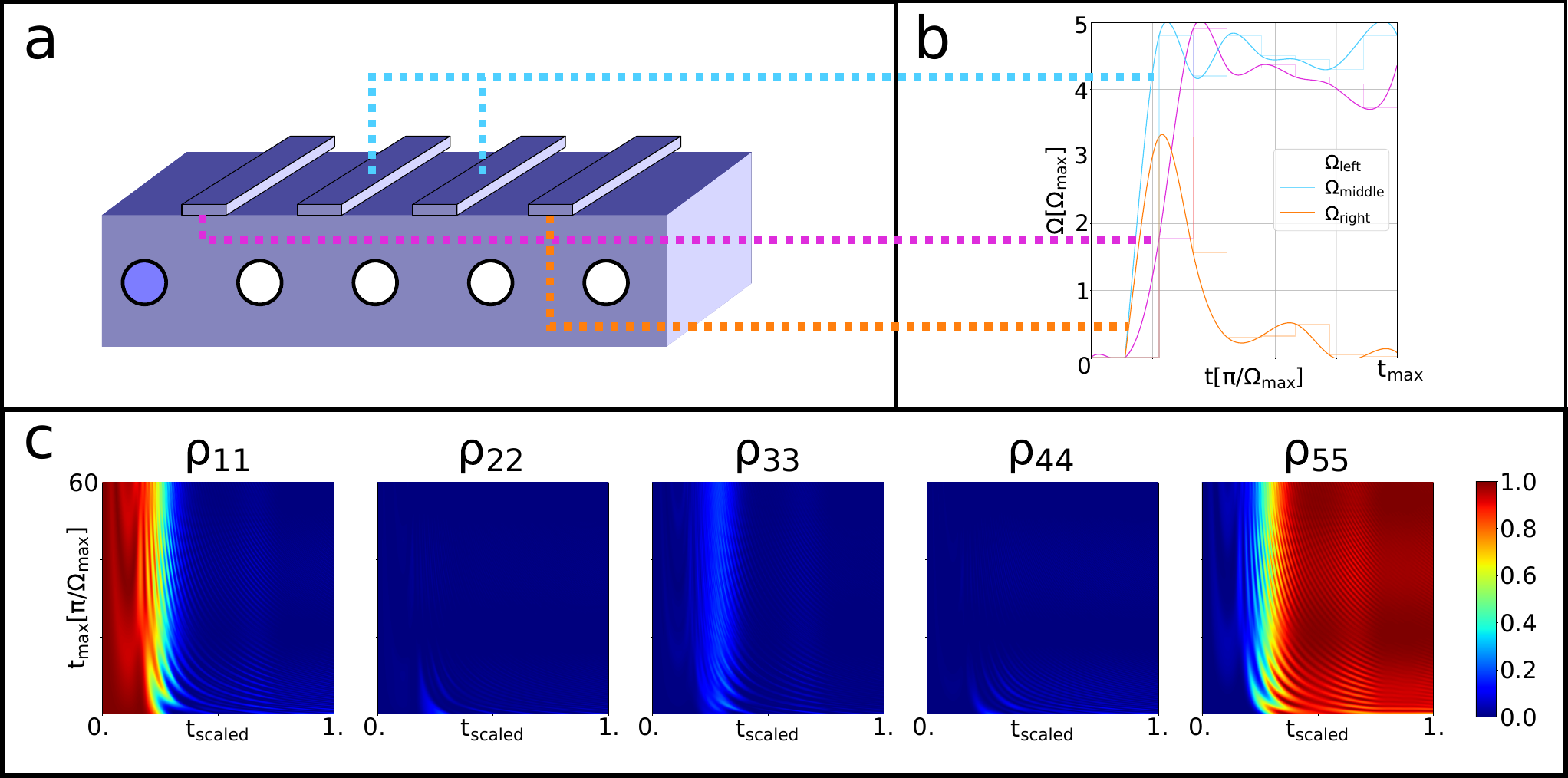}
	\caption{DRL-controlled Straggling CTAP (SCTAP) (a) Schematics of the five quantum dot array. The pair of central gates (b) are coupled according to Ref. \cite{greentree2004coherent}: $\Omega_{left}$ is tuned by the first coupling control gate, $\Omega_{right}$ by the last coupling control gate, while $\Omega_{middle}$ is identical as from the second and the third coupling control gates. In the example for $t_{max}=201 \frac{\pi}{\Omega_{max}}$. The dashed-thick pulses are discovered by the DRL, while the solid lines are cubic spline interpolations. (c) Population transfer by DRL-controlled SCTAP (Straggling CTAP). On the x-axis there is the (rescaled) time in unity of $t_{max}$, while on the y-axis there is $t_{max}$. During a time evolution, the increase/decrease of the electronic populations shows oscillations, which are reduced for some values of $t_{max}$, e.g. $t_{max}=21 \frac{\pi}{\Omega_{max}}$. 
For $t_{max}=21 \frac{\pi}{\Omega_{max}}$, the maximum value of $\rho_{33}$ is $\rho_{33,max}=0.1946$, while $\rho_{55,max}=0.99963689$, obtained with the fitted pulses. Notice how $\rho_{22}$, $\rho_{33}$ and $\rho_{44}$ are minimized during all the time evolution.
}
	\label{Fig4}
\end{figure}
\begin{figure}
	\centering
	\includegraphics[width=0.95\textwidth]{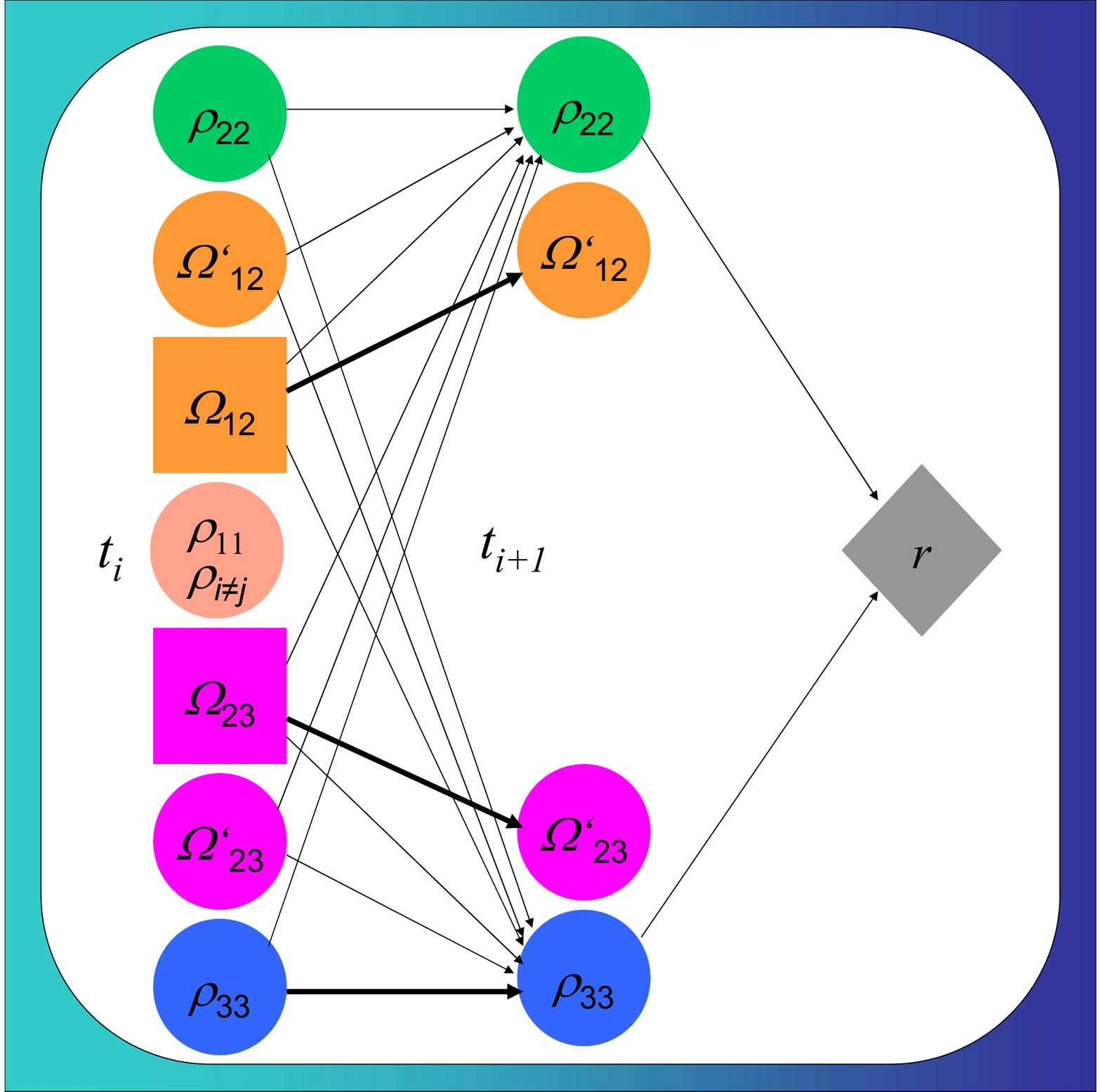}
	\caption{2-TBN for the ideal standard CTAP. The Grey diamond node is used to represent the reward function, the circles represent state variables and the squares represent action variables. The nodes are arranged on three vertical layers: the first layer on the left lists the variables at time $t_i$, the second layer the state variables at time $t_{i+1}$, and the third layer consists of the node that represents the reward function. Prime index indicates the value of a variable as stored at the previous step. A direct edge between two nodes means that the value of the source node influences the value of the target node. The thicker the line the greater the influence (values in the Supplementary Information). As expected, $\rho_{11}$ (shaded red) has no connections being linearly dependent on $\rho_{22}$ and $\rho_{33}$.}
	\label{Fig5}
\end{figure}

\cleardoublepage
\newpage

\bibliography{\jobname} 
\bibliographystyle{ieeetr}

\end{document}